# Illustrating field emission theory by using Lauritsen plots of transmission probability and barrier strength

Richard G. Forbes[a), b)]

*Advanced Technology Institute & Department of Electronic Engineering, Faculty of Engineering and Physical Sciences, University of Surrey, Guildford, Surrey GU2 7XH, UK*

Jonathan H. B. Deane

*Department of Mathematics, Faculty of Engineering and Physical Sciences, University of Surrey, Guildford, Surrey GU2 7XH, UK*

Andreas Fischer[c)], Marwan S. Mousa

*Department of Physics, Mu'tah University, Al-Karak 61710, Jordan*

───────────────────

[a)]American Vacuum Society member.

[b)]Electronic mail (permanent alias): r.forbes@trinity.cantab.net

[c)]Present address: Institut für Physik, Technische Universität Chemnitz, Germany.

**[Abstract]**

This technical note relates to the theory of cold field electron emission (CFE). It starts by suggesting that, to emphasize common properties in relation to CFE theory, the term "Lauritsen plot" could be used to describe all graphical plots made with the reciprocal of barrier field (or the reciprocal of a quantity proportional to barrier field) on the horizontal axis. It then argues that Lauritsen plots related to barrier strength ($G$) and transmission probability ($D$) could play a useful role in discussion of CFE theory. Such plots would supplement conventional Fowler-Nordheim (FN) plots. All these plots would be regarded as particular types of Lauritsen plot. The Lauritsen plots of $-G$ and $\ln D$ can be used to illustrate how basic aspects of FN tunnelling theory are influenced by the mathematical form of the tunnelling barrier. These, in turn, influence local emission current density and emission current. Illustrative applications used in this note relate to the well-known exact triangular and Schottky-Nordheim barriers, and to the Coulomb barrier (i.e., the electrostatic component of the electron potential energy barrier outside a model spherical emitter). For the Coulomb barrier, a good analytical series approximation has been found for the barrier-form correction factor; this can be used to predict the existence (and to some extent the properties) of related curvature in FN plots.

## I. INTRODUCTION

Fowler-Nordheim-type (FN-type) equations are a family of approximate equations that describe cold field electron emission (CFE) from a metal conduction band, for emitters that are "not too sharp" (tip radius of order 10 nm, or greater). FN-type equations are also used as empirical fitting equations for all types of CFE, but the interpretation of extracted results may be problematic if the emission situation is not orthodox[1-3].

The FN-type equations most suited to discuss basic theory give the local emission current density $J_L$ in terms of the local work-function $\phi$ and the local barrier field $F_L$. As is well known, for a planar emitter, a FN plot of type [$\ln\{J_L/F_L^2\}$ vs $F_L^{-1}$] is predicted to be nearly straight; for this reason, FN plots are widely used to represent and interpret experimental CFE data. One can also make theoretical plots of the form [$\ln D_F$ vs $F_L^{-1}$] and/or [$-G_F$ vs $F_L^{-1}$], where $D_F$ and $G_F$ are the tunneling probability and barrier strength defined below. This technical note suggests that such plots can be a useful (and perhaps under-appreciated) way of illustrating theoretical effects that relate specifically to the barrier form and the tunneling process, rather than partly to the summation over electron states (and, for currents, the integration over the emitting surface) used to derive FN-type equations.

The note's structure is as follows. Section II provides background theory; Section III illustrates the use of Lauritsen plots relating to tunneling probability and barrier strength, by describing five applications; and Section IV provides discussion. The usual electron emission convention is followed that fields, current densities and related quantities are treated as positive, even though negative in classical electromagnetism. In particular, the basic symbol $F$ denotes a positive quantity that is the negative of electrostatic field as used in classical electromagnetism.

To help discussion, this note introduces a special name for the class of data plots where the horizontal axis shows the reciprocal of either local barrier field $F_L$ or a parameter proportional to $F_L$. This plotting method was first used by Lauritsen[4-6], and was the critical breakthrough that led Fowler and Nordheim to develop CFE theory[7,8]. Thus, we call all plots of this kind "Lauritsen plots". The term "Millikan-Lauritsen (ML) plot", introduced in Ref. 9, describes the specific type of Lauritsen plot in which $\ln\{J_L\}$ (or a logarithm of a related quantity, such as emission current $i$) is plotted against $F_\mathrm{L}^{-1}$ or a linearly related quantity. FN plots, ML plots and Lauritsen plots of $\ln\{D_\mathrm{F}\}$ and $(-G_\mathrm{F})$ are all specific types of Lauritsen plot.

## II. Background theory

A technically complete FN-type equation for local emission current density $J_\mathrm{L}$ can be written in the abstract form

$$J_\mathrm{L} = Z_\mathrm{F} D_\mathrm{F} = Z_\mathrm{F} P_\mathrm{F} \exp[-G_\mathrm{F}], \tag{1}$$

where $D_\mathrm{F}$ is the tunnelling probability for a Fermi-level emitter electron moving "forwards" (normal to the emitter surface). This electron is said to be "in state F", and sees a barrier of zero-field height $\phi$. The subscript "F" is used to label quantities associated with this electron state or the related tunnelling barrier. $Z_\mathrm{F}$ is the "effective supply" for state F (i.e., the effective incident current density, for electrons approaching the emitter surface from the inside), and is found by summation over all occupied electron states.

Form (1) was originally introduced, in Refs 10 and 11, for free-electron models. For such models $Z_\mathrm{F}$ is easily calculated, in particular for planar or large-radius emitters[10], by the evaluation of appropriate integrals. However, because both $D_\mathrm{F}$ and $J_\mathrm{L}$ can be defined at all emitter surface positions, form (1) in fact applies at any surface position on an emitter of any shape, made from any material. The form of $Z_\mathrm{F}$ will depend on the circumstances, and may sometimes be very difficult to calculate.

A technically complete FN-type equation for the emission current $i$ can then be written

$$i = A_\mathrm{n} J_\mathrm{C}, \tag{2}$$

where $J_\mathrm{C}$ is a characteristic value of $J_\mathrm{L}$, and $A_\mathrm{n}$ is the related notional emission area[12].

As Eq. (1) shows, for deep tunneling the probability $D_\mathrm{F}$ can be written in the Landau and Lifschitz form[11,13] $D_F \approx P_F \exp[-G_F]$, where $G_F$ quantifies the barrier faced by an electron in state F, and $P_F$ is the related tunneling pre-factor. The parameter $G$ is defined by a JWKB-type (Jeffreys-Kramer-Wentzel-Brillouin-type) integration (e.g., Ref. 11):

$$G \equiv g_e \int M^{1/2}(z) dz ,  \qquad (3)$$

where $g_e$ [$\approx 10.24624$ eV$^{-1/2}$ nm$^{-1}$] is a universal constant called[14,15] the "JWKB constant for an electron". $M(z)$ is the motive energy that determines electron motion, and is given by $M=U-E_n$, where (in the one-dimensional Schrödinger equation) $U$ is the potential energy and $E_n$ is the total-energy component in the $z$-direction. The integral is taken "across the barrier", i.e., over the range of $z$ where $M(z) \geq 0$. For tunneling, $G$ can also be identified with the quantity $2|K|$ used by Fröman and Fröman[16].

In past literature, $G$ has been called the "Gamow exponent", the "WKB exponent" and the "JWKB exponent". We now prefer the physically more descriptive name "barrier strength"; thus $G_F$ is the barrier strength for state F. For consistency, we call Eq. (3) the "barrier-strength integral".

For an arbitrary, well-behaved, "general barrier" (GB), $G_F$ and $Z_F$ can be written in the forms

$$G_F^{GB} = \nu_F^{GB} G_F^{ET} = \nu_F^{GB} b \phi^{3/2}/F_L ,  \qquad (4)$$

$$Z_F^{GB} = \lambda_L^{GB} Z_F^{el} = \lambda_L^{GB} a \phi^{-1} F_L^2 ,  \qquad (5)$$

where $a$ and $b$ are the usual universal FN constants[15], $G_F^{ET}$ [$=b\phi^{3/2}/F_L$] is the barrier strength for the exact triangular (ET) barrier used in deriving the elementary FN-type equation[1,10], and $Z_F^{el}$ [$=a\phi^{-1}F_L^2$] is the effective-electron-supply term used[10] in the elementary equation. $\nu_F^{GB}$ is a correction factor related to the mathematical form of the general barrier, and $\lambda_L^{GB}$ is a local pre-exponential correction factor related to electron supply and summation over electron states; these factors correct the exponent and pre-exponential, respectively, of the elementary equation.

For the general barrier, writing Eq. (1) in so-called FN coordinates, yields

$$\ln\{J_L/Z_F^{GB}\} = \ln D_F^{GB} = \ln P_F^{GB} - G_F^{GB}, \tag{6}$$

The term $\ln P_F^{GB}$ is usually in the range $\{-1 < \ln P_F^{GB} < 1\}$ and is slowly varying; thus, barrier effects on FN plots are mainly due to the barrier-strength term $-G_F^{GB}$.

## III. APPLICATIONS

### A. The barrier strengths of the exact triangular and Schottky-Nordheim barriers

The first application compares the strengths of the two most commonly used barrier models: the exact triangular (ET) barrier, which has $M^{ET} = \phi - eF_L z$; and the Schottky-Nordheim (SN) barrier, which has $M^{SN} = \phi - eF_L z - e^2/16\pi\varepsilon_0 z$, where $e$ is the elementary positive charge, $\varepsilon_0$ is the electric constant, and $z$ is distance measured from the emitter's electrical surface. This comparison has been made before[17], but is presented differently here. It uses the scaled barrier field $f$ and the parameter $\eta$ (or $\eta^{SN}$) defined in Ref. 17 and also in Ref. 3.

In scaled form, the barrier strength $G_F^{ET}$ for an ET barrier of height $\phi$ is

$$G_F^{ET} = \eta/f . \tag{7}$$

The barrier strength $G_F^{SN}$ for an SN barrier of zero-field height $\phi$ can be written

$$G_F^{SN} = \eta \cdot v(f)/f \approx -\eta[f^{-1} - 1 - \tfrac{1}{6}\ln(f^{-1})] = -G_F^{ET} + \eta + (\eta/6)\ln(f^{-1}), \tag{8}$$

where the simple good expansion[17] $v(f) \approx 1 - f + \frac{1}{6} f \ln f$ has been used for the principal SN barrier function[17] $v(f)$. Hence the difference $\Delta(-G_F)$ between the strengths of the SN and ET barriers is

$$\Delta(-G_F) = [-G_F^{SN} - (-G_F^{ET})] \approx \eta + (\eta/6)\ln(f^{-1}). \tag{9}$$

In Fig. 1, the straight line ET shows how $-G_F^{ET}$ varies with $f^{-1}$, and the slightly curved line SN shows this for $-G_F^{SN}$. Point "R" is the reference point (1,0) at which curve SN starts, and line PL is a straight line drawn parallel to line ET, a vertical distance $\eta$ above it. Line PL passes through point R. Figure 1 shows that, for values of $f^{-1}$ of interest to practical field electron emission (approximately $2 \leq f^{-1} \leq 7$), much the larger contribution to $\Delta(-G_F)$ comes from the first term ($\eta$) on the right-hand-side of Eq. (9), i.e., from the constant upwards shift.

## B. Slope and intercept correction functions for the SN barrier

Figure 2 relates to the SN barrier and illustrates graphically, for $\phi$=4.50 eV and the specific value $f^{-1}$=5, the relationships between the barrier-form correction function $v(f)$, the slope correction function[17] $s(f)$ and the intercept correction function $r_{2012}(\eta, f)$. The function $r_{2012}(\eta, f)$ is a new type of intercept correction function introduced in Ref. 2 and given mathematically by

$$\ln\{r_{2012}(\eta, f)\} = \{s(f) - v(f)\}G_F^{ET}. \tag{10}$$

The value $f^{-1}$= 5 corresponds to $f$=0.2, and—for $\phi$= 4.50 eV ($\eta \approx 4.637$)—a barrier field $F_L \approx 2.8$ V/nm. For these values, Table I in Ref. 3 shows that $r_{2012} \approx 164$; hence, $\ln\{r_{2012}\} \approx 5.04$.

In the tangent method[2,18] of analyzing FN plots, an experimental data plot is modelled by the tangent to the theoretical equation, when the latter is written in FN coordinates. When, as in orthodox data analysis, an SN-barrier based FN-type equation is used to model the emission, the full equation for local emission current density $J_L^{SN}$ can (following the pattern of Eq. (6)) be written formally as

$$\ln\{J_L^{SN}/Z_F^{SN} P_F^{SN}\} = -G_F^{SN} = -v_F G_F^{ET}, \tag{11}$$

with

$$Z_F^{SN} = \lambda_L^{SN} a\phi^{-1} F_L^2, \tag{12}$$

where $\lambda_L^{SN}$ is the local pre-exponential correction factor for the SN barrier. A merit of the new function $r_{2012}$ is that it allows the equation for the tangent to Eq. (11) to be written

$$\ln\{J_L^{tan}/Z_F^{SN} P_F^{SN}\} = \ln\{r_{2012}\} - sb\phi^{3/2}/F_L = \ln\{r_{2012}\} - sG_F^{ET}. \tag{13}$$

Figure 2 illustrates graphically how the definitions fit together, when $G_F^{ET}$ corresponds to the value $f=5$. The slope of line "ET" is $-\eta$, and the slope of curve "SN" is (by definition) $-s(f)\eta$. Line V represents eq. (11), has slope $-v(f)\eta$, and gives the value of $-G_F^{SN}$ at point P; line T(5) represents Eq. (13) and gives its intersection with the axis, at $\ln\{r_{2012}\}$. This graphical definition of $r_{2012}$ is consistent with Eq. (10).

## C. Plot curvature in the deep tunneling regime

This application compares plot curvatures for the SN barrier and for the "Coulomb barrier" ($M^{CL}$) defined by the electrostatic component of the electron potential energy outside a sphere. For an emitter of radius $R$ and a barrier of height $\phi$, $M^{CL}$ is given by

$$M^{CL} = \phi - eF_L R(1 - R/r), \tag{14}$$

where $r$ is distance from the sphere centre, and the barrier field $F_L$ is defined as the field at $r=R$.

Edgcombe[19] has evaluated the related barrier-strength integral. In terms of a dimensionless parameter $\upsilon$ ("upsilon") (his $x_U$), called here "Edgcombe's parameter" and defined by

$$\upsilon = \phi / eF_L R, \tag{15}$$

his result can be written

$$-G_F^{CL} = -\nu_F^{CL} G_F^{ET} = -G_F^{ET} \times \frac{3}{2\upsilon}\left[\frac{\arcsin(\upsilon^{1/2})}{\{\upsilon(1-\upsilon)\}^{1/2}} - 1\right], \tag{16}$$

where $\nu_F^{CL}$ is the barrier-form correction factor for the Coulomb barrier.

This Coulomb-barrier tunneling problem was first addressed by Gamow[20] in 1928 (after the work of FN), in the context of explaining the Geiger-Nuttall law for $\alpha$-particle emission from nuclei. Various expressions related to Eq. (16) exist in the literature (e.g., in Ref. 13, §50, problem 2). Using the mathematical package MAPLE, we have checked that Eq. (16) is one of a number of equivalent mathematically correct forms.

Taylor expansion of $G_F^{CL}$ about $\upsilon=0$, using MAPLE, yields (after some algebraic manipulation)

$$\nu_F^{CL} = 1 + \tfrac{4}{5}\upsilon + \tfrac{24}{35}\upsilon^2 + \tfrac{64}{105}\upsilon^3 + O(\upsilon^4), \tag{17}$$

where the symbol $O(\upsilon^4)$ stands for terms of order $\upsilon^4$ and higher. In the range $0 \leq \upsilon \leq 0.5$, this expansion has an accuracy of better than 4%. As far as we are aware, this is the first time that this particular series expansion has been reported.

Obviously, in the limit of very large model radius $R$ (very small $\upsilon$) this result goes over into the result ($v_F^{ET}$=1) for the exact triangular barrier, which represents the electrostatic component of the electron potential energy outside a planar emitter.

In Eq. (17), the leading "unity" would generate a straight line (representing $-G_F^{ET}$) when $-G_F$ is plotted against $v$ or (for constant $\phi$ and $R$) against $F_L^{-1}$. The remaining terms cause a plot of $-G_F^{CL}$ versus $v$ or $F_L^{-1}$ to diverge downwards from this straight line, with the divergence and the curvature getting greater as $v$ or $F_L^{-1}$ increases, and (in the second case) with the effect being greater for smaller values of the model radius $R$. These effects are demonstrated in Figs. 3 and 4 which are Lauritsen plots that use the exact result (16) to plot $-v_F^{CL}$ against $v$, and $-G_F^{CL}$ against $F_L^{-1}$ for several $R$-values.

Unlike the case of the SN barrier (Fig. 1), where (a) the exact plot diverges upwards from line "PL" (and hence from line "ET") and (b) curvature of the exact plot is so small as to be hardly detectable, the plot curvature associated with a Coulomb barrier becomes noticeable when the model radius $R$ gets below a value between 20 and 50 nm. Another difference between the two cases is that, for the SN barrier, the curvature gets greater towards the left-hand side of the plot (the high-field, low-barrier-strength side), whereas for the Coulomb barrier the curvature gets greater towards the right-hand side (the high-barrier-strength side).

We cut off Lauritsen barrier-strength plots for $G_F$ values greater than 30, on the grounds that, in most experiments, the corresponding currents would be too small to detect.

These results provide useful qualitative understanding of effects likely to occur with small-radius real emitters, but caution is needed in applying them quantitatively. This is because the spherical-emitter model tends to lose its validity as $R$ becomes smaller than about 10 to 20 nm, due (a) to the influence of the emitter shank on the barrier form; and (b) to the possible onset of quantum confinement effects[21]. For the more realistic emitter shapes, with a shank and quasi-spherical end-region, the potential distribution associated with a sphere is no longer an adequate approximation to the solution of Laplace's equation for the more realistic emitter, when $R$ drops below about 10 to 20 nm. This point has previously been made by Edgcombe[22].

Another point is that formula (17) performs quantitatively well only if $F_L > 2\phi/eR$; however, this is often not a practical difficulty. For $\phi$= 4.5 eV and $R$=10 nm, this condition implies that $F_L$ should be greater than about 1 V/nm. Thus, for emitter radii where the model is physically adequate, formula

(17) should usually be mathematically adequate for $F_L$-values of practical interest.

**D. Breakdown of the planar-emitter deep-tunneling approximations**

For both the ET and SN barriers, the well-known formulas for tunneling probability $D_F$ are approximations valid in the deep-tunneling transmission regime, i.e., when the barrier strength is positive and sufficiently large.

For the ET barrier, there exists an exact general formula for $D_F$ that is known to be mathematically correct[14,23]. This has the mathematical form[14]

$$D_F^{ET} = 1/[\tfrac{1}{2} + \tfrac{1}{4}\pi\omega_F(A^2 + B^2) + \tfrac{1}{4}\pi\omega_F^{-1}(A'^2 + B'^2)], \qquad \text{(FD)} \qquad (18)$$

where $A$, $B$, $A'$ and $B'$ are values of the Airy functions and their derivatives, evaluated at a defined (field-dependent) value of their argument, and $\omega_F$ is a parameter that depends on the barrier field $F_L$ and on the Fermi energy (i.e., the kinetic energy $K_F$ of an electron in state F). Equation (18) is a special case (for barrier height $\phi$) of Eq. (2.19) in Ref. 14, where fuller mathematical details can be found.

Expression (18) applies both to electron tunneling and to wave-mechanical electron transmission over the top of the barrier, termed "flyover" in Ref. 14. Thus, the $D_F^{ET}$ that appears in Eq. (18) is a probability for transmission across the barrier, whether this transmission takes place by tunnelling or by flyover. Tunneling probability is special form of transmission probability, so we use the more general name for $D_F^{ET}$ in what follows.

For tunneling, the well-known original FN approximate formula[7] for $D_F^{ET}$ is

$$D_F^{ET} \approx P_F^{FN} \exp(-G_F^{ET}) \equiv [4K_F^{1/2}\phi^{1/2}/(K_F + \phi)] \cdot \exp(-G_F^{ET}), \qquad \text{(FN)} \qquad (19)$$

where FN's tunneling pre-factor $P_F^{FN}$ is defined by the term in square brackets in Eq. (19). This formula is an asymptotic approximation to Eq. (18), valid for deep tunneling.

For a barrier of height $\phi$=4.50 eV, Figs. 5(a) and (b) are Lauritsen plots of $\ln(D_F^{ET})$ that compare the results of evaluating expressions (19) (marked "FN") and (18) (marked "FD") as functions of barrier field; Fig. 5(a) shows in more detail the curve behaviour at very high fields (very low values of $F_L^{-1}$). Since values of $D_F$>1 ($\ln D_F$>0) are unphysical, Fig. 5(a) shows clearly that the FN approximate result (19) becomes unphysical if extrapolated to sufficiently high fields.

The exact and approximate curves for $\ln(D_F^{ET})$ begin to diverge for barrier strengths less than about 3 to 5. Such barriers occur at fields higher than those normally used in CFE, and the barriers are weaker than those normally used, which most commonly have strengths inside the range 5 to 30. The limiting behaviour at high fields is thus of small practical relevance, but is of interest for the theory of tunneling. At extremely high fields (extremely low values of $1/F_L$) the exact curve goes through a maximum and tends towards $-\infty$. This mathematical behaviour was first noticed by Rokhlenko[24]. For $\phi$= 4.50 eV, the maximum occurs at a barrier field of around 850 V/nm, i.e., far beyond any value that can be realized experimentally.

In the case of the SN barrier, it is mathematically impossible to obtain an exact analytical solution of the Schrödinger equation in terms of the established functions of mathematical physics. However, there exists a semi-classical approximation formula for transmission probability $D_F$ that is considered to be mathematically reliable over most (perhaps all) fields of practical interest to CFE. This was derived from the work of Fröman and Fröman[16] and can be put in the mathematical form

$$D_F = P_F \exp(-G_F) / [1 + P_F \exp(-G_F)] = 1/[1 + (P_F)^{-1} \exp(G_F)]. \qquad (20)$$

For most barriers, exact analytical expressions for $P_F$ in terms of the established functions of mathematical physics are not known and probably do not exist. Mayer has developed[25-27] numerical procedures able to calculate $P_F$. However, to evaluate transmission probability $D_F^{SN}$ for the SN barrier, it is more common to use the Kemble approximation[28,29] (later derived in a different mathematical way by Miller and Good[30]). In effect, this sets $P_F$=1 in Eq. (20) and (for the SN barrier) takes

$$D_F^{SN} \approx 1/[1+\exp(G_F^{SN})].  \qquad \text{(Kem)} \qquad (21)$$

For the SN barrier, wave-mechanical flyover (transmission over the barrier) corresponds to the scaled-barrier-field range $1 < f \leq \infty$. The principal SN barrier function $v(f)$ is continuous through the value $f=1$, and remains well-defined[32,33] for $f>1$ (although the approximation given earlier for $v(f)$ increasingly loses its accuracy above $f=1$, and performs very poorly above about $f=4$). As $f$ increases above 1, $v(f)$ becomes increasingly negative[32]. It follows that the barrier strength $G_F^{SN}$ remains mathematically well-defined for $f>1$, but becomes increasingly negative as $f$ increases. In the high-field limit where $f \to \infty$, it is found that $v(f) \to -\infty$, $G_F^{SN} \to -\infty$, and formula (21) evaluates to the physically correct limit $D_F^{SN} \to 1$.

For sufficiently large positive values of $G_F^{SN}$ (i.e., for deep tunneling), Eq. (21) reduces to the simple-JWKB formula

$$D_F^{SN} \approx \exp(-G_F^{SN}) . \qquad \text{(JWKB)} \qquad (22)$$

Expression (25) can, of course, be mathematically evaluated for all values of $G_F^{SN}$.

For the SN barrier, Figs 5(c) and (d) are Lauritsen plots that compare the behavior of the Kemble approximation (marked "Kem") with that of the simple-JWKB approximation (marked "JWKB"). In behaviour analogous to that of the ET barrier, the simple-JWKB approximation clearly becomes unphysical when extrapolated to very high fields. Obviously, this is because, as field increases and $G_F^{SN}$ decreases, expression (22) goes outside its regime of mathematical validity. The more-accurate Kemble approximation diverges from the simple-JWKB approximation for barrier strengths less than around 4.

As already noted, this divergence occurs outside the range of field and barrier-strength values normally of practical interest to cold field electron emission. However, the difference in behavior between the approximations is of relevance in Schottky emission, where the emitter is heated and significant numbers of electrons tunnel through weak barriers.

**E. Comparison of theoretical high-field (low barrier-strength) behaviors**

For transmission theory, it is of theoretical interest to compare the high-field limiting behaviors of Eq. (18) (the exact ET-barrier treatment) and Eq. (21) (the Kemble approximation for the SN barrier). The main difference, as illustrated in Figs 5(a) and (c), is that for the ET barrier the transmission probability $D_F \to 0$, but for the Kemble approximation $D_F \to 1$. This difference can be explained qualitatively: in the ET case, the electron always has to tunnel through a barrier, which gets increasingly "spiky" at high fields; in the SN case, the barrier is pulled down below the Fermi level at fields greater than a reference field $F_R$ (about 14 V/nm for a $\phi$=4.5 eV emitter). Above this field the situation becomes one of wave-mechanical flyover: the electron transmission is over the top of the barrier and $D_F$ is expected to become unity as the "height (in energy)" above the barrier top increases.

A more fundamental difference between Eqs. (18) and (21) is that there are three terms in the denominator of Eq. (18) but only two terms in the denominator of Eq. (21). The basic mathematical reason is clear: the three-term formula is generated by wave-matching[14] at a sharp break in slope at the apex of the triangular barrier, whereas the two-term formula is derived by continuous integrations[16] along a smooth path in complex space that stays well clear of the mathematical zeroes associated with the barrier.

**IV. Discussion**

This technical note has given five illustrations of the use of Lauritsen plots, mainly of the natural logarithm of the transmission probability $D$ or of (the negative of) the barrier strength $G$. We emphasize that these plots are not FN plots as conventionally understood, because they do not explicitly involve a current or a current density.

To make these Lauritsen plots directly relevant to FN-type equations, we have used plots that relate to barriers with zero-field height equal to the local work function $\phi$. However, these Lauritsen plots could equally well be used to illustrate the characteristics of barriers of any (positive) height.

These Lauritsen plots illustrate, in a convenient way, the direct effects of barrier form (i.e., the form of $M(z)$) on barrier strength and transmission probability. Via Eqs (1) and (2), they also indicate how barrier and transmission effects influence field dependences in emission current density and current. (In simple models, FN plots are parallel to the corresponding Lauritsen plot of $\ln D_F$; in more realistic models, additional curvature will often be introduced.)

For practical field electron emission, the most interesting results are probably those related to plot curvature for a spherical emitter. The Coulomb-barrier results here clearly illustrate that one possible cause of curvature in conventional FN plots is a barrier-form effect: if the field fall-off across the width of a real emitter barrier is sufficiently large, then a corresponding conventional FN plot is expected to be curved, particularly so at the low-field (high barrier strength) side of the plot.

Because an algebraic series expansion has been found for the barrier-form correction factor $v_F^{CL}$ for the Coulomb barrier, there seems a reasonable chance that future work may be able to develop satisfactory algebraic expressions for the related slope and (new-type) intercept correction factors, although (due to the onset of shank effects) these would not be accurate for practical emitters of apex radius less than around 20 nm.

More generally, comparisons can be made between a Lauritsen plot of $\ln\{D_F\}$ and the corresponding FN plot (for example, a plot involving $\ln\{J_M/F_M^2\}$, where $F_M$ is macroscopic field and $J_M$ is macroscopic current density). Differences in shape between the plots would suggest/confirm that factors other than transmission probability were affecting the experimental plot shape, and that additional investigation might be helpful. Detailed discussion of applications of this kind is beyond the scope of the present paper.

In conclusion, we believe that the illustrations above of the uses of Lauritsen plots of $\ln D$ and $-G$ and have demonstrated the merits of these plots in discussions of CFE theory. We also believe that other forms of Lauritsen plot, such as those used to display the properties of the SN barrier functions in Figs (2c) and (2d) of Ref. 17, will prove useful in future work. A long-term potential advantage is that these Lauritsen plots would retain their usefulness, even if community practice eventually moved away[9] from the use of FN plots to interpret CFE experimental data.

## ACKNOWLEDGMENTS

Andreas Fischer thanks the Alexander von Humboldt foundation for a Feodor Lynen fellowship and Mu'tah University for hospitality.

**Figure 1**

Fig. 1. Comparison of barrier-strength dependences on inverse scaled barrier field, for the exact triangular (ET) and Schottky-Nordheim (SN) barrier models. Line PL is drawn parallel to line ET, a distance $\eta$ above it. Curve SN starts at the reference point "R", at (1,0).

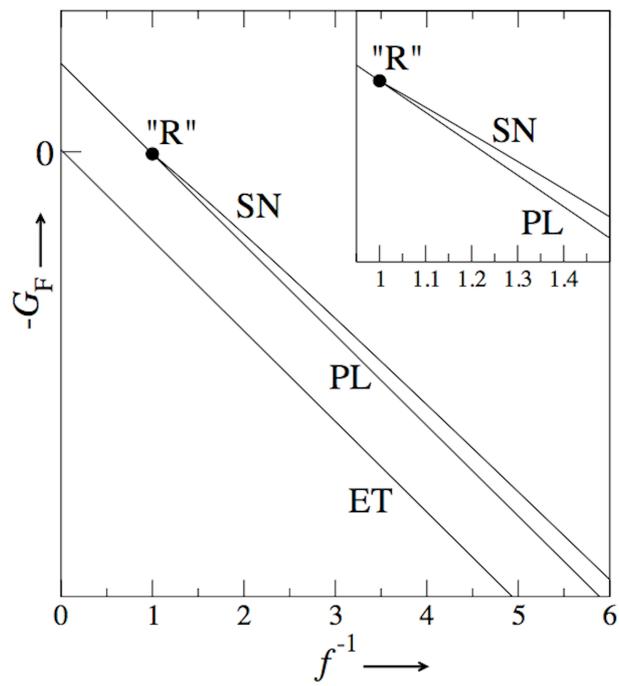

# Figure 2

Fig. 2. To illustrate the relationships between the SN-barrier correction functions $v(f)$, $s(f)$ and $r_{2012}(\phi,f)$, for the specific values $\phi=4.50$ eV ($\eta \approx 4.637$), $f=0.2$. The line T(5) is the tangent to curve SN at point "P", at which $f^{-1}=5$. The slopes of lines ET, V and T(5) are, respectively, $-\eta$, $-\eta \cdot v(0.2)$ and $-\eta \cdot s(0.2)$, and $G_F^{ET}=5\eta$.

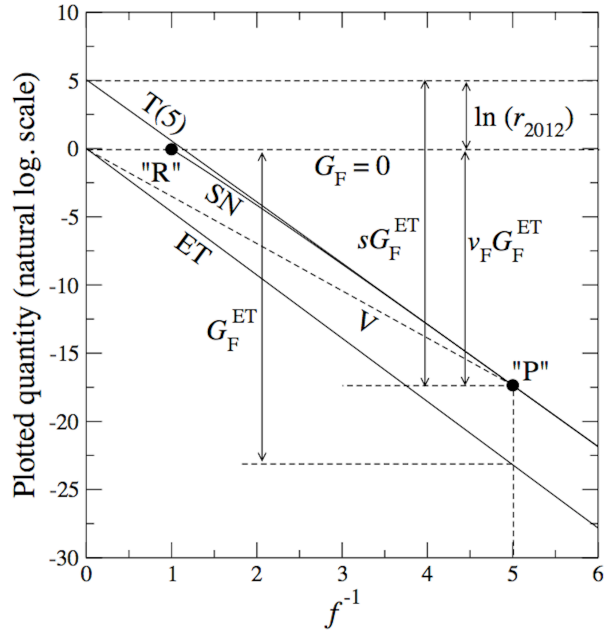

**Figure 3**

Fig. 3. To show how the barrier-form correction factor for the Coulomb barrier ($v_F^{CL}$) varies with Edgcombe's parameter $\upsilon$ ("upsilon"), defined by Eq. (18).

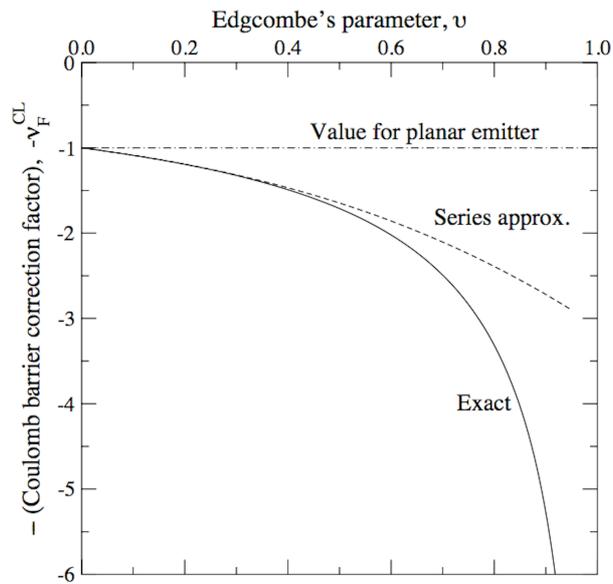

**Figure 4**

Fig. 4. To show how, for a Coulomb barrier, the barrier strength for state F varies with inverse barrier field, for the work-function value 4.50 eV, and the emitter radii shown. For sufficiently small model radii, the curvature in the Lauritsen plot is detectable.

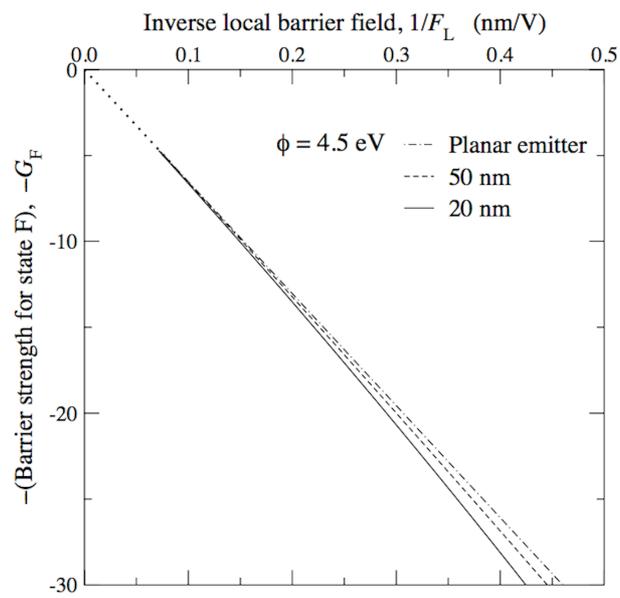

**Figure 5**

Fig. 5. To show how the transmission probability $D_F$ varies with inverse barrier field. Figures 5(a) and (b) show results for an exact triangular barrier of height 4.50 eV, as predicted by the original Fowler-Nordheim formula (FN) and by an exact treatment (FD). Figures 5(c) and (d) show results for a Schottky-Nordheim barrier of zero-field height 4.50 eV, as predicted by the usual simple-JWKB treatment (JWKB) and by the Kemble approximation (Kem). For each barrier, the left-hand figure shows the high-field (low $F_L^{-1}$) region in greater detail.

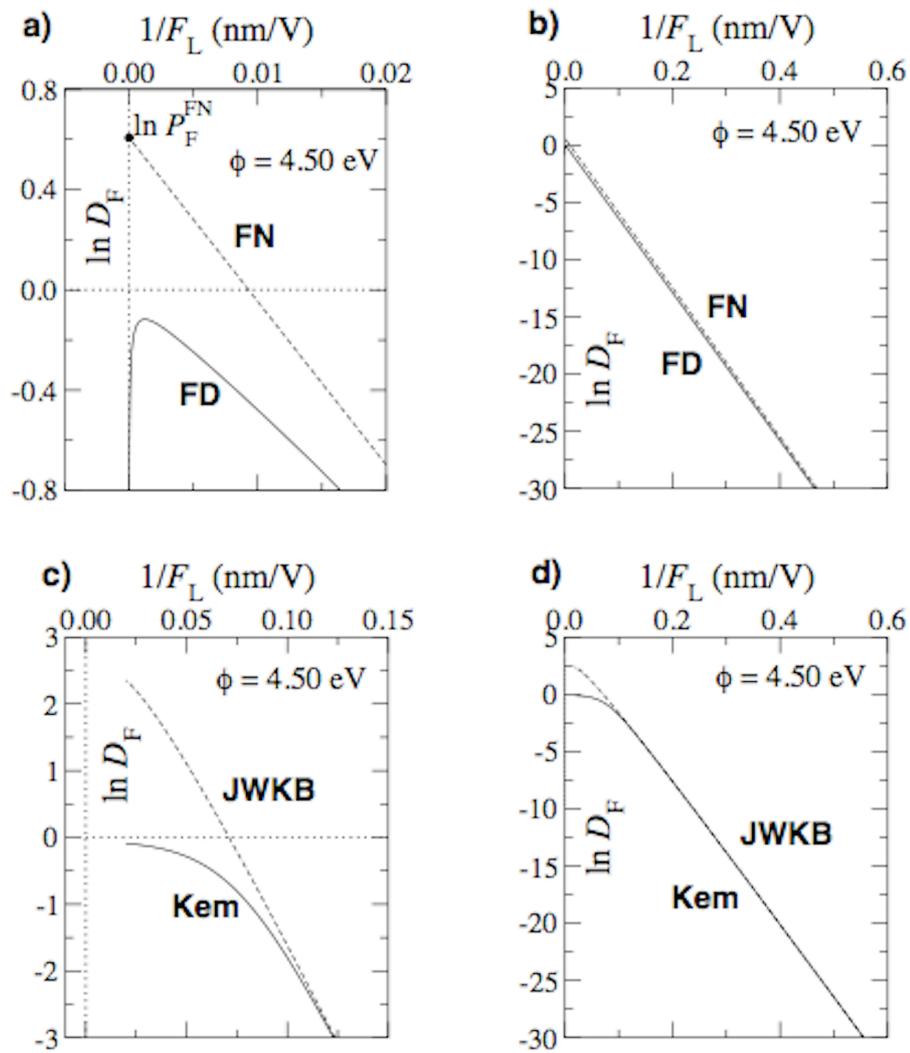